\documentclass[a4paper,12pt]{article}

\usepackage[body={18cm, 25cm, centered}]{geometry}

\usepackage[utf8]{inputenc}

\begin{document}

\title{On the possible experimental revelations of Unruh and Sokolov-Ternov effects}
\date{}

%\maketitle

	\begin{center}
		\baselineskip=16pt
		{\Large \bf 
		    On the possible experimental revelations of Unruh and Sokolov-Ternov effects
		}
		\vskip 1cm
            Emil T.\ Akhmedov$^{a,b,}$\footnote{\texttt{akhmedov@itep.ru}},
			Kirill Gubarev$^{a,b,}$\footnote{\texttt{kirill.gubarev@phystech.edu}}
		\vskip .3cm
		\begin{small}
			{\it 
			    $^a$Moscow Institute of Physics and Technology, 141700, Dolgoprudny, Russia,\\
			    $^b$National Research Centre “Kurchatov Institute”, 123182, Moscow, Russia
			}
		\end{small}
	\end{center}
\begin{center} 
\textbf{Abstract}
\end{center} 
\begin{quote}
In this paper we propose generalizations of the Sokolov-Ternov and Unruh effects, and discuss the possibility to measure them on different experiments.
\end{quote}

\vspace{5mm}

{\bf 1.} The possibility to heat up a physical system by just accelerating it, as is predicted in \cite{Unruh:1976db}, is closely related to the phenomenon of the black hole radiation \cite{Hawking:1975vcx} (see e.g. \cite{Birrell:1982ix} for a review). This Unruh effect can be understood as due to the presence of specific correlations between zero-point fluctuations along the world-line of the non-inertially moving physical system (e.g. detector\footnote{The crucial difference of the Hawking effect is that there the radiation is essentially present everywhere in space-time, but is seen in different manner in from different reference systems.}). Namely, the probability per unit time to excite the detector is proportional to \cite{Birrell:1982ix}:

\begin{equation}
    w \propto \int_{-\infty}^{+\infty} d\tau \, e^{- \frac{i}{\hbar} \tau\, \Delta E}\, G\left[\underline{x}\left(t-\frac{\tau}{2}\right), \, \underline{x}\left(t+\frac{\tau}{2}\right)\right],\label{prob}
\end{equation}
where $\underline{x}(\tau) \equiv x_\mu(\tau)$, $\mu = 0,\dots,d$ is the world-line of the detector, $\Delta E > 0$ is the excitation energy of the detector's internal levels and $G\left[\underline{x}_1, \, \underline{x}_2\right] = \left\langle \phi(\underline{x}_1) \, \phi(\underline{x}_2) \right\rangle$ is the two-point Wightman function of the scalar field $\phi(x)$, if we are considering the detector which is coupled to (interacting with) the scalar field. If the detector is coupled to a gauge or other type of fields (even massive) the expression for the probability rate is similar.

The effect is interesting when the motion of the detector is stationary. In such a case, the Wightman correlator 

\begin{equation}
G\left[\underline{x}\left(t-\frac{\tau}{2}\right), \, \underline{x}\left(t+\frac{\tau}{2}\right)\right] = F(\tau), \label{Ftau}    
\end{equation}
is just a function of the proper time difference, $\tau$, between the two points on the world-line of the detector. Then the expression (\ref{prob}) does describe the probability rate far enough from the  start of the acceleration, i.e. when all the radiation and absorption processes of the scalar field by the detector are stationarized. To achieve the stationarity we have to take the time of observation $t$ much larger than the time of the detector equilibration. 

Being formulated in such a way the effect can be present even for acceleration for finite time and has nothing to do with the presence of the horizon in the eternally accelerating reference frame. In such a case the number of the detected particles is an invariant (similar in spirit to the proper time), which depends only on the world line $x(\tau)$ of the detector between the moments of start of observation, $\tau_1$, and its end, $\tau_2$.

The Unruh effect is predicted on the basis of analytic properties of the function $F(\tau)$ from (\ref{Ftau}) in the complex $\tau$-plane. In fact, the non zero contribution to the integral (\ref{prob}) is gained from the poles of $F(\tau)$ in the lower complex $\tau$-plane. The poles appear from the singularity of the Wightman function in the vicinity of the light-cone, which is well known and universal for any type of the field:

\begin{equation}
    G\left[\underline{x}, \, \underline{y}\right] \propto \frac{1}{(x_0 - y_0 - i \,\epsilon)^2 - \left|\vec{x} - \vec{y}\right|^2}. 
\end{equation}
(The coefficient of proportionality here is important to predict the equilibration time, but that is an irrelevant technical detail for our considerations. One just has to keep in mind that for orbiting motion the equilibration time is much shorter than for the linear acceleration.)
Then for the case of constant linear acceleration, $a$, along the first axis 

$$
\underline{x}(\tau) = \frac{c^2}{a} \, \Big(\sinh \left(\frac{a\tau}{c}\right), \cosh\left(\frac{a\tau}{c}\right), 0, \dots, 0\Big),
$$
one finds that

$$
F(\tau) \propto \frac{a^2}{c^2 \, \sinh^2\left[\frac{a}{2c}\, (\tau - i \, \epsilon)\right]}.
$$
This function has poles in the lower half of the complex $\tau$-plane, whose regular positions and residues lead to the well known exact result:

$$
w \propto \frac{\Delta E}{e^{\frac{2\pi c \Delta E}{\hbar a}} - 1},
$$
i.e. one finds the seminal thermal distribution with the Unruh temperature $T = \hbar a/(2\pi c k_B)$, where $k_B$ is the Boltzmann constant. This rate is closely related to the particle number which can be found via quantization of the field theory in the Rindler's non-inertial reference frame accompanying the eternally accelerating observer \cite{Birrell:1982ix}.

Being understood in such a way the Unruh effect is closely related to the so called Sokolov-Ternov effect \cite{Sokolov:1963zn} (see e.g. \cite{LL4} for a review). The presence of the Unruh effect within the Sokolov-Ternov one was discussed in \cite{Bell:1982qr} and  \cite{Bell:1986ir}, while the equivalence between these two effects was established in \cite{Akhmedov:2006nd} and 
\cite{Akhmedov:2007xu}.

The Sokolov-Ternov effect is observed experimentally. Its experimental revelation is that ultrarelativistic electron beams in accelerator circular rings are not completely polarized. The point is that certain part of the synchrotron radiation comes from the flip of the electron spin, i.e. goes with the change of the spin energy levels in the external magnetic field. The ultrarelativistic spin is playing the role of the two-level detector\footnote{One has to consider ultrarelativistic particles to neglect the Landau levels and consider the particle's world-line as classical.}, which is coupled to the electromagnetic field rather than to a scalar one. However, the flip rate is still given by an expression, which is similar to the one in (\ref{prob}) with the presence of certain preexponential factors under the integral \cite{Akhmedov:2006nd} (see also \cite{LL4}).

In all, as is explained in \cite{LL4}, the Sokolov-Ternov effect also appears due to the pole structure of the Wightman function $F(\tau)$ from (\ref{Ftau}) for the electromagnetic field. I.e. this effect also appears due to the correlation of the zero-point fluctuations along the world-line corresponding to the circular motion. For orbiting motion with the radius $R$ and angular velocity $\omega$ the world-line is 

$$\
\underline{x}(\tau) = \Big(c \gamma\, \tau, \, R\, \cos(\gamma \omega \tau), R\, \sin(\gamma \omega \tau), 0\Big),
$$ 
where $\gamma = 1/\sqrt{1 - \omega^2 \, R^2 / c^2}$ and $\omega R = v$. Then

$$
F(\tau) \propto \frac{1}{\left[c \gamma\, \tau - i \, \epsilon\right]^2 - 4\, R^2 \, \sin^2\left(\frac{\gamma \, \omega \, \tau}{2}\right)}.
$$
This function also has poles in the lower half of the complex $\tau$-plane, but their positions can be found only by solving the transcendental equation of the kind $\tau = A \, \sin(B\tau)$ for some constants $A$ and $B$. Hence, the poles are impossible to find analytically. Nevertheless taking into account the closest pole to the real axis in the lower half-plane gives:

$$
w \propto a \, e^{- \sqrt{12} \, \frac{c \Delta E}{\hbar a}}, \quad {\rm where} \quad a = \gamma^2 \, \omega^2 \, R = \frac{\gamma^2 v^2}{R}.
$$
The problem is that for the ultrarelativistic electron spin in constant magnetic fields $\sqrt{12} \, \frac{c \Delta E}{\hbar a} \approx 1/\gamma + (g-2)/2$, where $g\approx 2$ for electron. As the result, the exponential factor in the Sokolov-Ternov effect is almost one and the whole effect (eight percent of depolarized electrons in a homogeneous magnetic field) comes from the preexponential in the electromagnetic analog of (\ref{prob}) (see \cite{LL4} and \cite{Akhmedov:2006nd}).

In this paper we propose generalizations of the Sokolov-Ternov effect in which the exponential factor plays a crucial role\footnote{See also \cite{Galon:2020lol} for recent discussion of other kinds of generalizations of the Sokolov-Ternov effect for the electron and positron beams in storage-rings.}. We hope that some day these effects will be measured bringing further understanding and evidence in favour of the Unruh effect for the linear acceleration and of the Hawking effect.

{\bf 2.} Consider a beam of protons $p$ performing a circular motion in an accelerator of the length $L=2\pi R$. Based on the observations made above, we predict that, when the equilibrium has been reached, some fraction of the total number $n_{p}$ of protons $p$ in the beam will be in an excited state $N^+$ due to the orbital acceleration $a$. The fraction can be estimated as

\begin{eqnarray}
    P_{p\rightarrow N^{+}} \propto n _{p} \, \exp\bigg(- \frac{\sqrt{12} c \, \Delta E_{p\rightarrow N^{+}}}{a \hbar}\bigg) = \nonumber \\ = n_{p} \,  \exp\bigg(- \frac{\sqrt{12} c L \Delta {E}_{p\rightarrow N^{+}}}{v^2 h \gamma^{2}}\bigg)  \, \stackrel{v\approx c}{\approx} \, n_{p} \,  \exp\bigg(- \frac{\sqrt{12} L \Delta {E}_{p\rightarrow N^{+}}}{c h \gamma^{2}}\bigg).
\end{eqnarray}
The question is if the acceleration $a$ and the flux $n_p$ are large enough for this effect to be seen, because the energy difference $\Delta E_{p\rightarrow N^{+}}$ for the case under consideration is really big. But below we will consider other proposals where the energy difference is much smaller. Thus, let us consider concrete examples where the effect can be potentially seen.

\begin{itemize}

\item \underline{\textbf{CERN}} Explicit values for the CERN are: the length is $L = 27 \cdot 10^{3} m$, the number of particles in the beam $n_{p} \sim 10^{12}$, the proton energy in the beam is $E_{p} \sim 7 \cdot 10^{12} eV$, that corresponds to $\gamma \sim 7.5 \cdot 10^{3}$, the energy gap is $\Delta E_{p\rightarrow N^{+}} \sim 100 \cdot 10^{6} eV = 10^{8} eV$, and $ch = 4 \cdot 10^{-15} eV \cdot s \cdot 3 \cdot 10^{8} \frac{m}{s} = 10^{-6} eV \cdot m$. Then

\begin{eqnarray}
        P_{p\rightarrow N^{+}} \propto n _{p} \, \exp\bigg(- \frac{\sqrt{12} L \Delta E_{p\rightarrow N^{+}}}{c h \gamma^{2}}\bigg) = \nonumber \\ = 10^{12} \, \exp\bigg(- \frac{3.5 \cdot 27 \cdot 10^{3} m \cdot 10^{8} eV}{(7.5 \cdot 10^{3})^2 \cdot 10^{-6} eV \cdot m}\bigg) \sim 10^{12} \, \exp(- 10^{11}),    
\end{eqnarray}
which obviously makes the effects hopeless to measure in the present conditions. We can find the effective temperature of the Sokolov--Ternov effect for the CERN experiment
\begin{equation}
    k_{B} T^{CERN}_{ST} \sim \frac{10^8 \, eV}{10^{11}} \approx 10^{-3} eV \sim 10 K.
\end{equation}

%...............................

%that means the degree of exponent

%\begin{equation}
 %   30 \simeq \frac{\pi R \cdot 10^{15}}{\gamma^2 \cdot 1 m} \quad \Rightarrow \quad \frac{R \cdot 10^{15}}{\gamma^2 \cdot 1 m} \simeq 10.
%\end{equation}

%For $R = 100m$ then $\gamma \simeq 10^{8}$, that means $\mathcal{E}_{p} = 10^6 TeV$.

\item \underline{\textbf{NICA}} However, may be there is a hope to 
measure the effect in the transitions $ion\rightarrow ion^{*}$, where the energy difference is substentially smaller?
Namely, we propose to consider a different analog of Unruh-Sokolov-Ternov effect, where the role of the detectors play ions in circular accelerator rings with their internal energy levels. Consider e.g. ions circulating in NICA experiment. The length of the NICA collider is $L = 5 \cdot 10^{2} m$ with the number of ions in the beam being equal to $n_{ions} \sim 10^{10}$. The ion's energy in the beam $E_{ion} \sim 5 \cdot 10^{9} \frac{eV}{nucleon}$, which corresponds to $\gamma \sim 5$. The energy difference between the ground and excited states of ion being equal to $\Delta E_{ion\rightarrow ion^{*}} \sim 10^{6} eV$. Then

\begin{eqnarray}
    P_{ion\rightarrow ion^{*}} \propto n_{ion} \, \exp\bigg(- \frac{\sqrt{12}L \Delta E_{ion\rightarrow ion^{*}}}{c h \gamma^{2}}\bigg) = \nonumber \\ = 10^{10} \, \exp\bigg(- \frac{3.5 \cdot 5 \cdot 10^{2} m \cdot 10^{6} eV}{(5)^2 \cdot 10^{-6} eV \cdot m}\bigg) \sim 10^{10} \, \exp(- 10^{14}),
\end{eqnarray}
which makes the situation even worth than above. The effective temperature of the Sokolov--Ternov effect for the NICA experiment is
\begin{equation}
    k_{B} T^{NICA}_{ST} \sim \frac{10^6 \, eV}{10^{14}} \approx 10^{-8} eV \sim 10^{-4} K.
\end{equation}

\item However, the situation can be made a bit better if one could consider not completely ionized atoms in the orbiting beams. Namely consider hydrogen type atoms (with only one electron) orbiting e,g, in the NICA type of accelerator. Then the electron's energy levels will be with the approximate difference $\Delta E_{atom\rightarrow atom^{*}} \sim 10 eV$, which will change the rate to:

\begin{equation}
    P_{atom\rightarrow atom^{*}} \sim 10^{10} \, \exp(- 10^{9}).
\end{equation}
This is still hopeless to measure.

\item We can reformulate the problem in a different way. How big should be the flux $n_{p}$ to have at least one atom in the beam to get excited? The answer is obviously $n_{atom} \sim e^{10^9}$ for the last case. Too humongous to be achievable at present time even though luminosity of accelerators is increasing rather rapidly. 

May be then one can hope to see the effect on the hyperfine structure of the atomic energy levels, where the energy difference is $10^{-6}$ smaller than between the regular atomic levels. That makes $n_{atom} \sim e^{10^3}$. However, in this case the main mistake for the effect that we consider will come from the fact that particles in the beam are thermalized during non-homogeneous acceleration periods and are not in the vacuum state. Even though for the NICA experiment the effective temperature of Sokolov--Ternov effect is tiny $T^{NICA}_{ST} \sim 10^{-4} K$, for the CERN experiment it is $T^{CERN}_{ST} \sim 10 K$, that can be of the same order as the temperature of in the beam. Furthermore, the Sokolov--Ternov effect may play role in the crystalline ion beams \cite{SCHRAMM2004583} where the temperature is of order $T \sim 10^{-3} K$.

%E.g. if we could achieve somehow such values as the acllerator ring of the radius $R=10^{3} m$ with proton energy in the ring $E_{p} = 100 TeV$ that corresponds to $\gamma \simeq 10^{5}$ then for the effect to be seen the number of proton in the beam should be $n_{p} \simeq exp(10^8).$ At the same time the expected number $n_{p} \simeq 10^{14}$ in the nearest future. Again with such an energy difference as in the transition $p\rightarrow N^{+}$ the effect is practically hopeless to measure.

\end{itemize} 

{\bf 3.} We would like to thank Pavel Pakhlov for valuable comments. This work is supported by the Foundation for the Advancement of Theoretical Physics and Mathematics “BASIS” grant, by RFBR grant 19-02-00815 and joint RFBR-
MOST grant 21-52-52004 MNT a, and by Russian Ministry of education and science.

\end{document}